# Bound states of spatially-dependent mass Dirac equation with the Eckart potential including Coulomb tensor interaction

Sameer M. Ikhdair[1,2] and Babatunde J. Falaye [3*]

[1] *Department of Physics, Faculty of Science, An-Najah National University, New Campus, Nablus, West Bank, Palestine*

[2] *Department of Physics, Near East University, 922022 Nicosia, Northern Cyprus, Mersin 10, Turkey*

[3] *Theoretical Physics Section, University of Ilorin, Ilorin, Nigeria*

[*]Email: fbjames11@physicist.net

## Abstract

We investigate the approximate solutions of the Dirac equation with the position-dependent mass particle in the Eckart potential field including the Coulomb tensor interaction by using the parametric Nikiforov-Uvarov method. Taking an appropriate approximation to deal with the centrifugal term, the Dirac energy states and the corresponding normalized two-spinor components of the wave function are obtained in closed form. Some special cases of our solution are investigated. Further, we present the correct solutions obtained via asymptotic iteration method which are in agreement with the parametric Nikiforov-Uvarov method results.

**Keywords:** Dirac equation; position-dependent mass; Eckart potential; Coulomb-like tensor potential; Nikiforov-Uvarov method

**PACS:** 03.65.Ge, 03.65.Fd, 03.65.Pm, 02.30.Gp

## 1. Introduction

Over the past years, the position-dependent mass (PDM) solutions of the non-relativistic and relativistic quantum systems have received much attention. Many authors have used different methods to study both exactly and quasi-exactly solvable Schrödinger, Klein-Gordon and Dirac equations in the presence of variable mass having suitable mass distribution function for different potentials (see [1-4] and references therein).

The study of relativistic effects is always useful in some quantum mechanical systems [5-8]. Therefore, the Dirac equation has become the most appealing relativistic wave equation for spin-1/2 particles. For example, in the relativistic treatment of nuclear phenomena the Dirac equation is used to describe the behavior of the nuclei in nucleus and also in solving many problems of high-energy physics and chemistry. For this reason, it has been used extensively to study the relativistic heavy ion collisions, heavy ion spectroscopy and more recently in laser–matter interaction (for a review, see [9] and references therein) and condensed matter physics [10].

On the other hand, the Dirac equation with PDM has been found to be very useful in studying some physical properties of various microstructures [11-18]. Recently, there has been increased interest in searching for analytical solutions of the Dirac equation with PDM and with constant mass under the spin and pseudospin (p-spin) symmetries [19-25]. In their recent works Hamzavi and Rajabi have studied Dirac equation in 1+1 dimensions with scalar-vector-pseudoscalar Cornell potential for spin-$1/2$ particle under spin and pseudospin symmetries [26]. The ground state energy of the KG equation with scalar-vector Cornell potentials [27] has been calculated under the influence of the magnetic and Aharonov-Bom fields.

Moreover, we studied the scalar charged particle exposed to relativistic scalar-vector Killingbeck (Cornell plus harmonic oscillator) potentials in presence of magnetic and Aharonov-Bohm flux fields and obtained its energy eigenvalues and wave functions using the analytical exact iteration method [28]. Hamzavi et al. have solved approximately the Dirac equation with the Eckart potential including the Coulomb-like tensor interaction, in view of pseudospin symmetry, using the Nikiforov-Uvarov method [29].

The aim of the present work is to obtain the energy eigenvalues and normalized two-components wave function in the PDM Dirac equation with the vector Eckart potential field and the Coulomb-like tensor interaction using the parametric Nikiforov-Uvarov (pNU) method [30-35] which has been used by many authors to solve the non-relativistic as well as relativistic wave equations for various potential models (see for example, [36-44] and references therein). We also present the correct Dirac energy equation for the Eckart potential within the asymptotic iteration method (AIM).

In the non-relativistic level, approximate solution of the Schrödinger equation for the Eckart potential and parity-time symmetric version including centrifugal term has been obtained to investigate the energy eigenvalues and wave functions for various values of $n$ and $l$ using the Nikiforov-Uvarov (NU) method [45]. A new proper approximation scheme to the centrifugal term was applied to study the Schrödinger equation with an Eckart-like potential and the bound and scattering state energy levels with normalization constant of the wave function were obtained [46-48]. The analytical approximation to the $l$-wave solutions of the Eckart potential have been obtained in the tridiagonal representation [49].

The Eckart potential is defined as [50,51]

$$V(r)=V_1 \cos ech^2\left(\alpha r\right)-V_2 \coth\left(\alpha r\right), \tag{1}$$

and can be rewritten in the exponential form as

$$V(r)=A\frac{e^{-2\alpha r}}{\left(1-e^{-2\alpha r}\right)^2}-B\frac{1+e^{-2\alpha r}}{1-e^{-2\alpha r}}, \quad S(r)=0, \quad (0<2\alpha r<\infty,\ A;B>0,\ A<B) \tag{2}$$

where the parameters $A=4V_1$ and $B=V_2$ describe the depth of the potential well, both $A$ and $B$ are positive and real parameters; $\alpha$ is a positive parameter to control the width of the potential well having the dimension of the inverse length. The potential is an asymmetric function. The interesting results concerning the potential give inspiration not only to explore another similar potential, but also to study thermal reaction of Formaldehyde [52]. It has been widely applied in physics [53] and chemical physics [54,55]. In Figure 1, we show the behavior of the Eckart potential as a function of $r$ for three screening parameter values $\alpha=0.2,\ 0.5,\ 0.8\ fm^{-1}$ by taking the strength parameters $V_1=V_2=0.8.$ It is seen that the potential strength decreases with the increasing of the screening parameter value.

This paper is organized as follows. In Section 2, we briefly introduce the PDM Dirac equation with scalar and vector Eckart potentials including Coulomb-like tensor interaction under spin and p-spin symmetric limits for any arbitrary spin-orbit coupling quantum number $\kappa$-state. The approximate Dirac energy eigenvalue equation and the corresponding normalized spinor wave function are also obtained by taking an appropriate approximation to the centrifugal term $\sim r^{-2}$. In Section 3, we consider a few special cases of our solution like Hulthén potential, generalized Morse potential, nonrelativistic Eckart potential and the correct Dirac energy states within AIM. Finally, we give our conclusion in Section 4.

## 2. Position-dependent mass Dirac equation with a tensor coupling potential

The Dirac equation for a spin-$1/2$ single particle with a position-dependent mass $m(r)$ moving in the field of an attractive scalar potential $S\left(r\right)$, a repulsive vector potential $V\left(r\right)$ and a tensor potential $U\left(r\right)$ is given by (in units $\hbar=c=1$) [56]

$$\left[\vec{\alpha}.\vec{p}+\beta\left(m(r)+S\left(r\right)\right)-i\beta\vec{\alpha}.\hat{r}U\left(r\right)\right]\psi\left(\vec{r}\right)=\left[E-V\left(r\right)\right]\psi\left(\vec{r}\right), \tag{3}$$

where $\vec{\alpha}$ and $\beta$ are the $4\times4$ usual Dirac matrices and $\vec{p}$ is the three-dimensional momentum operator given by

$$\vec{\alpha}=\begin{pmatrix}0 & \vec{\sigma}\\ \vec{\sigma} & 0\end{pmatrix},\quad \beta=\begin{pmatrix}I & 0\\ 0 & -I\end{pmatrix},\quad \vec{p}=-i\vec{\nabla} \tag{4}$$

respectively, with $I$ is $2\times2$ unitary matrix and $\vec{\sigma}$ are three-vector spin matrices

$$\sigma_1=\begin{pmatrix}0 & 1\\ 1 & 0\end{pmatrix},\quad \sigma_2=\begin{pmatrix}0 & -i\\ i & 0\end{pmatrix},\quad \sigma_3=\begin{pmatrix}1 & 0\\ 0 & -1\end{pmatrix}. \tag{5}$$

Further, $E$ is the total relativistic binding energy of the system, and $m(r)$ is the position-dependent mass of the fermionic particle. The Dirac eigenfunction can be written as follows

$$\psi_{n\kappa}\left(\vec{r}\right)=\begin{pmatrix}f_{n\kappa}\left(\vec{r}\right)\\ g_{n\kappa}\left(\vec{r}\right)\end{pmatrix}=\frac{1}{r}\begin{pmatrix}F_{n\kappa}\left(r\right)Y_{jm}^{l}\left(\theta,\varphi\right)\\ iG_{n\kappa}\left(r\right)Y_{jm}^{\tilde{l}}\left(\theta,\varphi\right)\end{pmatrix}, \tag{6}$$

where $f_{n\kappa}\left(\vec{r}\right)$ and $g_{n\kappa}\left(\vec{r}\right)$ are the radial functions for the upper (large) and the lower (small) components, respectively, $Y_{jm}^{l}\left(\theta,\varphi\right)$ and $Y_{jm}^{\tilde{l}}\left(\theta,\varphi\right)$ are spin and pseudo spin (p-spin) spherical harmonics, respectively, $n$ is the radial quantum number and $m$ is the projection of the angular momentum on the $z$-axis. The orbital angular momentum quantum numbers $l$ and $\tilde{l}$ stand for the spin and p-spin quantum numbers, respectively. For a given spin-orbit coupling term $\kappa=0,\pm1,\pm2,\ldots,$ the total angular momentum, the orbital angular momentum and the pseudo orbital angular momentum are given by $j=|\kappa|-1/2,$ $l=|\kappa+1/2|-1/2$ and $\tilde{l}=|\kappa-1/2|-1/2,$ respectively.

By substituting Eqs. (6) and (4) into Eq. (3) and making use of the relations

$$\left(\vec{\sigma}.\vec{A}\right)\left(\vec{\sigma}.\vec{B}\right)=\vec{A}.\vec{B}+i\,\vec{\sigma}.\left(\vec{A}\times\vec{B}\right), \tag{7a}$$

$$\left(\vec{\sigma}.\vec{P}\right)=\vec{\sigma}.\hat{r}\left(\hat{r}.\vec{P}+i\,\frac{\vec{\sigma}.\vec{L}}{r}\right), \tag{7b}$$

together with the following properties

$$\begin{aligned}
&\left(\vec{\sigma}.\vec{L}\right)Y_{jm}^{\tilde{l}}\left(\theta,\phi\right)=\left(\kappa-1\right)Y_{jm}^{\tilde{l}}\left(\theta,\phi\right),\\
&\left(\vec{\sigma}.\vec{L}\right)Y_{jm}^{l}\left(\theta,\phi\right)=-\left(\kappa+1\right)Y_{jm}^{l}\left(\theta,\phi\right),\\
&\left(\vec{\sigma}.\hat{r}\right)Y_{jm}^{\tilde{l}}\left(\theta,\phi\right)=-Y_{jm}^{l}\left(\theta,\phi\right),\\
&\left(\vec{\sigma}.\hat{r}\right)Y_{jm}^{l}\left(\theta,\phi\right)=-Y_{jm}^{\tilde{l}}\left(\theta,\phi\right),
\end{aligned} \tag{8}$$

the Dirac equation reduces to the following coupled differential equations whose solutions are the upper and lower radial wave functions $F_{n\kappa}\left(r\right)$ and $G_{n\kappa}\left(r\right)$,

$$\left(\frac{d}{dr}+\frac{\kappa}{r}-U\left(r\right)\right)F_{n\kappa}\left(r\right)=\left(m(r)+E_{n\kappa}-\Delta\left(r\right)\right)G_{n\kappa}\left(r\right), \tag{9a}$$

$$\left(\frac{d}{dr}-\frac{\kappa}{r}+U\left(r\right)\right)G_{n\kappa}\left(r\right)=\left(m(r)-E_{n\kappa}+\Sigma\left(r\right)\right)F_{n\kappa}\left(r\right), \tag{9b}$$

where $\Sigma\left(r\right)=V\left(r\right)+S\left(r\right)$ and $\Delta\left(r\right)=V\left(r\right)-S\left(r\right)$ are the sum and difference potentials, respectively. Eliminating $F_{n\kappa}\left(r\right)$ and $G_{n\kappa}\left(r\right)$ from Eqs. (9a) and (9b), we can obtain the following two second-order differential equations for the upper and lower radial spinor components:

$$\begin{aligned}
&\left[\frac{d^2}{dr^2}-\frac{\kappa\left(\kappa+1\right)}{r^2}+\frac{2\kappa}{r}U\left(r\right)-\frac{dU\left(r\right)}{dr}-U^2\left(r\right)\right.\\
&-\left(m(r)+E_{n\kappa}-\Delta\left(r\right)\right)\left(m(r)-E_{n\kappa}+\Sigma\left(r\right)\right)\\
&\left.-\frac{\left(\frac{dm\left(r\right)}{dr}-\frac{d\Delta\left(r\right)}{dr}\right)}{m(r)+E_{n\kappa}-\Delta\left(r\right)}\left(\frac{d}{dr}+\frac{\kappa}{r}-U\left(r\right)\right)\right]F_{n\kappa}\left(r\right)=0,
\end{aligned} \tag{10}$$

$$\begin{aligned}
&\left[\frac{d^2}{dr^2}-\frac{\kappa\left(\kappa-1\right)}{r^2}+\frac{2\kappa}{r}U\left(r\right)+\frac{dU\left(r\right)}{dr}-U^2\left(r\right)\right.\\
&-\left(m(r)+E_{n\kappa}-\Delta\left(r\right)\right)\left(m(r)-E_{n\kappa}+\Sigma\left(r\right)\right)\\
&\left.-\frac{\left(\frac{dm\left(r\right)}{dr}+\frac{d\Sigma\left(r\right)}{dr}\right)}{M-E_{n\kappa}+\Sigma\left(r\right)}\left(\frac{d}{dr}-\frac{\kappa}{r}+U\left(r\right)\right)\right]G_{n\kappa}\left(r\right)=0,
\end{aligned} \tag{11}$$

respectively, where $\kappa(\kappa+1)=l\,(l+1)$ and $\kappa(\kappa-1)=\tilde{l}\,(\tilde{l}+1)$. The spin-orbit quantum number $\kappa$ is related to the quantum numbers for spin symmetry $l$ and p-spin symmetry $\tilde{l}$ as

$$\kappa=\begin{cases} -(l+1)=-(j+1/2) & (s_{1/2},p_{3/2},\ldots,\text{etc.}) \quad j=l+1/2, \ \text{aligned spin}\,(\kappa<0) \\ +l=+(j+1/2) & (p_{1/2},d_{3/2},\ldots,\text{etc.}) \quad j=l-1/2, \ \text{unaligned spin}\,(\kappa>0), \end{cases} \tag{12}$$

and the quasidegenerate doublet structure can be expressed in terms of a p-spin angular momentum $\tilde{s}=1/2$ and pseudo-orbital angular momentum $\tilde{l}$, which is defined as

$$\kappa=\begin{cases} -\tilde{l}=-(j+1/2) & (s_{1/2},p_{3/2},\ldots,\text{etc.}) \quad j=\tilde{l}-1/2, \ \text{aligned p-spin}\,(\kappa<0) \\ +(\tilde{l}+1)=+(j+1/2) & (d_{3/2},f_{5/2},\ldots,\text{etc.}) \quad j=\tilde{l}+1/2, \ \text{unaligned p-spin}\,(\kappa>0). \end{cases} \tag{13}$$

For example, $\left(1s_{1/2},0d_{3/2}\right)$ and $\left(1p_{3/2},0f_{5/2}\right)$ can be considered as p-spin doublets.

### 2.1. Energy eigenvalues and wave functions of the spin symmetric case

To obtain analytical solution to the Dirac equation for a single particle with PDM in the field of Eckart potential including the Coulomb-like tensor interaction, we need to approximate the spin-orbit centrifugal term. This can be achieved by using an approximation in the following form as [57,58]

$$\frac{1}{r^2}=\lim_{\alpha\to 0}\left[4\alpha^2\,\frac{e^{-2\alpha r}}{(1-e^{-2\alpha r})^2}\right]. \tag{14}$$

where $\alpha$ is the screening parameter describing the range of the Eckart potential which is valid only for $\alpha r\ll 1$. The tensor potential as Coulomb-like potential, that is,

$$U(r)=-\frac{T}{r}, \qquad T=\frac{Z_aZ_be^2}{4\pi\varepsilon_0}, \qquad r\geq R_c, \tag{15}$$

where $R_c=7.78fm$ is the Coulomb radius, $Z_a$ and $Z_b$ denote the charges of the projectile $a$ and the target nuclei $b$, respectively [59].

To solve Eq. (10) with Eqs. (2) and (14)-(16), we must eliminate the last term in Eq. (10) by letting $\frac{dm(r)}{dr}-\frac{d\Delta(r)}{dr}=0$ which gives the PDM function $m(r)-V(r)=m_0=\text{constant},$ or alternatively

$$m(r) = m_0 + 4V_1 \frac{e^{-2\alpha r}}{\left(1-e^{-2\alpha r}\right)^2} - V_2 \frac{1+e^{-2\alpha r}}{1-e^{-2\alpha r}}. \tag{16}$$

Substituting Eqs. (2) and (14)-(16) into Eq. (10), we obtain

$$\begin{aligned}&\left[\frac{d^2}{dr^2} - 4\alpha^2(\kappa+T)(\kappa+T+1)\frac{e^{-2\alpha r}}{\left(1-e^{-2\alpha r}\right)^2}\right.\\&\left.-\left(m_0+E_{n\kappa}\right)\left(m_0-E_{n\kappa}+8V_1\frac{e^{-2\alpha r}}{\left(1-e^{-2\alpha r}\right)^2}-2V_2\frac{1+e^{-2\alpha r}}{1-e^{-2\alpha r}}\right)\right]F_{n\kappa}(r)=0.\end{aligned} \tag{17}$$

Using the new variable $s = e^{-2\alpha r}$ where $s \in (0,1)$, the above equation becomes

$$\begin{aligned}&\left[\frac{d^2}{ds^2}+\frac{1}{s}\frac{d}{ds}-\left((\kappa+T)(\kappa+T+1)+\frac{2\left(m_0+E_{n\kappa}\right)V_1}{\alpha^2}\right)\frac{1}{s(1-s)^2}\right.\\&\left.-\frac{\left(m_0^2-E_{n\kappa}^2\right)}{4\alpha^2 s^2}+\frac{\left(m_0+E_{n\kappa}\right)V_2}{2\alpha^2}\frac{(1+s)}{s^2(1-s)}\right]F_{n\kappa}(s)=0,\end{aligned} \tag{18}$$

where the upper component satisfies the asymptotic behavior at the boundaries, i.e., $F_{n\kappa}(0) = F_{n\kappa}(1) = 0.$ In terms of the new parameters, we have

$$\left\{\frac{d^2}{ds^2}+\frac{1}{s}\frac{d}{ds}+\frac{1}{s^2(1-s)^2}\left[-\left(\xi_0+\xi_2\right)s^2+\left(2\xi_0-\xi_1\right)s-\left(\xi_0-\xi_2\right)\right]\right\}F_{n\kappa}(s)=0, \tag{19}$$

with

$$\xi_0=\frac{m_0^2-E_{n\kappa}^2}{4\alpha^2},\quad \xi_1=(\kappa+T)(\kappa+T+1)+\frac{2\left(m_0+E_{n\kappa}\right)V_1}{\alpha^2},\quad \xi_2=\frac{\left(m_0+E_{n\kappa}\right)V_2}{2\alpha^2}. \tag{20}$$

Equation (19) is convenient for obtaining the parametric NU solution. The outline of this method is given in Appendix A [30-35]. Now by comparing Eq. (19) with Eq. (A2), we can obtain the constants

$$\begin{aligned}&c_1=1, && p_2=\frac{\left(m_0+E_{n\kappa}\right)\left(m_0-E_{n\kappa}+2V_2\right)}{4\alpha^2},\\&c_2=1, && p_1=-(\kappa+T)(\kappa+T+1)+\frac{2\left(m_0+E_{n\kappa}\right)\left(m_0-E_{n\kappa}-4V_1\right)}{4\alpha^2},\\&c_3=1, && p_0=\frac{\left(m_0+E_{n\kappa}\right)\left(m_0-E_{n\kappa}-2V_2\right)}{4\alpha^2},\end{aligned} \tag{21}$$

and further using Eq. (A7), we have

$$c_4=0,\ \ c_5=-\frac{1}{2},\ \ c_6=\frac{1}{4}+\frac{\left(m_0+E_{n\kappa}\right)\left(m_0-E_{n\kappa}+2V_2\right)}{4\alpha^2},\tag{22a}$$

$$c_7=\left(\kappa+T\right)\left(\kappa+T+1\right)-\frac{2\left(m_0+E_{n\kappa}\right)\left(m_0-E_{n\kappa}-4V_1\right)}{4\alpha^2},\tag{22b}$$

$$c_8=\frac{\left(m_0+E_{n\kappa}\right)\left(m_0-E_{n\kappa}-2V_2\right)}{4\alpha^2},\ \ c_9=\left(\kappa+T+1/2\right)^2+\frac{8\left(m_0+E_{n\kappa}\right)V_1}{4\alpha^2},\tag{22c}$$

$$c_{10}=\sqrt{\frac{\left(m_0+E_{n\kappa}\right)\left(m_0-E_{n\kappa}-2V_2\right)}{\alpha^2}},\ \ c_{11}=\sqrt{\left(2\kappa+2T+1\right)^2+\frac{8\left(m_0+E_{n\kappa}\right)V_1}{\alpha^2}},\tag{22d}$$

$$c_{12}=\sqrt{\frac{\left(m_0+E_{n\kappa}\right)\left(m_0-E_{n\kappa}-2V_2\right)}{4\alpha^2}},\ c_{13}=\frac{1}{2}\left(1+\sqrt{\left(2\kappa+2T+1\right)^2+\frac{8\left(m_0+E_{n\kappa}\right)V_1}{\alpha^2}}\right).\tag{22e}$$

The energy equation can be obtained by using Eqs. (A5), (21) and (22) as

$$E_{n\kappa}^2=m_0^2-4\alpha^2\left[\frac{1}{4}\left(\frac{4\xi_2}{1+2n+\sqrt{1+4\xi_1}}-\frac{1+2n+\sqrt{1+4\xi_1}}{2}\right)^2+\xi_2\right],$$

which shows the energy dependence on the parameters $\xi_1$ and $\xi_2$ or alternatively it can be expressed in terms of the potential parameters, spin-orbit quantum number $\kappa$ and tensor coupling strength $T$ as

$$E_{n\kappa}^2=m_0^2-\frac{4\left(m_0+E_{n\kappa}\right)^2V_2^2}{\alpha^2\left(1+2n+\sqrt{\left(2\kappa+2T+1\right)^2+\frac{8\left(m_0+E_{n\kappa}\right)V_1}{\alpha^2}}\right)^2}-\frac{\alpha^2\left(1+2n+\sqrt{\left(2\kappa+2T+1\right)^2+\frac{8\left(m_0+E_{n\kappa}\right)V_1}{\alpha^2}}\right)^2}{4},\tag{23}$$

which is identical to Eq. (21) of Ref. [60] for the $s$-wave $(\kappa=-1)$ and the absence of tensor interaction $(T=0)$. It is also consistent with the result given by Eq. (77) of Ref. [61] in the Klein-Gordon (KG) theory when $\kappa\rightarrow l$ and $T=0.$ On the other hand, Eq. (23) is the energy equation of spin-$1/2$ particle in the Dirac theory with Eckart potential $\Sigma(r)=\Delta(r)=V(r)$ is identical but physically different with Eq. (22) of Ref. [62] which is the energy equation for a spin-$0$ particle in the KG theory with equal Eckart scalar and vector potentials. It is worthy to mention that the energy equation (23) is same as the energy equation of the constant mass Dirac equation with equally mixed scalar-vector Eckart potentials in the presence of exact spin symmetric limit, i.e., $\Sigma(r)=V(r)+S(r)=2V(r)$ and $\Delta(r)=V(r)-S(r)=C_s=0.$

In the limiting case when the screening parameter $\alpha\to 0$ (low screening regime), we have

$$\lim_{\alpha\to 0}E_{n\kappa}=m_0-\frac{1}{2V_1}\left(V_2^2+4V_1^2\right). \tag{24}$$

On the other hand, to find the corresponding wave functions, referring to Appendix relation (A6), we find the functions

$$\rho(s)=s^{2\sqrt{\xi_0-\xi_2}}(1-s)^{\sqrt{1+4\xi_1}}, \tag{25}$$

$$\phi\left(s\right)=s^{\sqrt{\xi_0-\xi_2}}(1-s)^{\frac{1}{2}\left(1+\sqrt{1+4\xi_1}\right)}. \tag{26}$$

Hence, relation (A6) with the help of the weight function (25) gives

$$y_n\left(s\right)=P_n^{\left(2\sqrt{\xi_0-\xi_2},\sqrt{1+4\xi_1}\right)}\left(1-2s\right). \tag{27}$$

Moreover, by using $F_{n\kappa}(s)=\phi(s)y_n(s)$, we can obtain the upper component of the Dirac spinor as

$$\begin{aligned}
&F_{n\kappa}(s)=Ns^{\sqrt{\xi_0-\xi_2}}(1-s)^{\left(1+\sqrt{1+4\xi_1}\right)/2}P_n^{\left(2\sqrt{\xi_0-\xi_2},\sqrt{1+4\xi_1}\right)}\left(1-2s\right),\\
&F_{n\kappa}(r)=\mathrm{N}'\left(e^{-2\alpha r}\right)^{\sqrt{\xi_0-\xi_2}}(1-e^{-2\alpha r})^{\left(1+\sqrt{1+4\xi_1}\right)/2}\\
&\times{}_2F_1\left(-n,n+2\sqrt{\xi_0-\xi_2}+\sqrt{1+4\xi_1}+1;2\sqrt{\xi_0-\xi_2}+1;e^{-2\alpha r}\right),\\
&\mathrm{N}'=\frac{\Gamma\left(n+2\sqrt{\xi_0-\xi_2}+1\right)}{\Gamma\left(2\sqrt{\xi_0-\xi_2}+1\right)n!}N,
\end{aligned} \tag{28}$$

where $N$ is the normalization constant. On the other hand, the lower component of the Dirac spinor can be calculated from Eq. (9a) as

$$\begin{aligned}
&G_{n\kappa}\left(r\right)=\frac{1}{\left(m_0+E_{n\kappa}\right)}\left\{\left[\alpha\left(1+\sqrt{1+4\xi_1}\right)\frac{e^{-2\alpha r}}{(1-e^{-2\alpha r})}-2\alpha\sqrt{\xi_0-\xi_2}+\frac{\kappa+T}{r}\right]F_{n\kappa}\left(r\right)\right.\\
&+\mathrm{N}'\left(e^{-2\alpha r}\right)^{1+\sqrt{\xi_0-\xi_2}}(1-e^{-2\alpha r})^{\left(1+\sqrt{1+4\xi_1}\right)/2}\frac{2\alpha n\left(n+2\sqrt{\xi_0-\xi_2}+\sqrt{1+4\xi_1}+1\right)}{\left(2\sqrt{\xi_0-\xi_2}+1\right)}\\
&\left.\times{}_2F_1\left(1-n,n+2\sqrt{\xi_0-\xi_2}+\sqrt{1+4\xi_1}+2;2\sqrt{\xi_0-\xi_2}+2;e^{-2\alpha r}\right)\right\},
\end{aligned} \tag{29}$$

where we have followed the derivation presented in Ref. [35].

### 2.2. Pseudospin symmetric limit

In order to normalize the wave function, we calculate $G_{n\kappa}(r)$. Following the same procedures of solution as before by inserting Eq. (2), (14), (15) and $m(r)=m_0-V(r)$ into Eq. (11), we can obtain

$$\left[\frac{d^2}{dr^2}-4\alpha^2\left(\kappa+T-1\right)\left(\kappa+T\right)\frac{e^{-2\alpha r}}{\left(1-e^{-2\alpha r}\right)^2}\right.$$
$$\left.-\left(m_0-E_{n\kappa}\right)\left(m_0+E_{n\kappa}-8V_1\frac{e^{-2\alpha r}}{\left(1-e^{-2\alpha r}\right)^2}+2V_2\frac{1+e^{-2\alpha r}}{1-e^{-2\alpha r}}\right)\right]G_{n\kappa}\left(r\right)=0. \tag{30}$$

Thus, making transformation of variables as before, we reduce Eq. (30) into the simple form:

$$\left(\frac{d^2}{ds^2}+\frac{1}{s}\frac{d}{ds}-\xi_0-\frac{\xi_3}{s\left(1-s\right)^2}+\frac{\left(1+s\right)\xi_4}{s^2\left(1-s\right)}\right)G_{n\kappa}\left(s\right)=0, \tag{31a}$$

$$\xi_3=\left(\left(\kappa+T\right)\left(\kappa+T-1\right)-\frac{2\left(m_0-E_{n\kappa}\right)V_1}{\alpha^2}\right),\quad \xi_4=-\frac{\left(m_0-E_{n\kappa}\right)V_2}{2\alpha^2}. \tag{31b}$$

To avoid repetition in the solution of Eq. (31a), we follow the same procedures explained in the subsection 2.1 and hence obtain the energy eigenvalue equation:

$$E_{n\kappa}^2=m_0^2-4\alpha^2\left[\frac{1}{4}\left(\frac{4\xi_4}{1+2n+\sqrt{1+4\xi_3}}-\frac{1+2n+\sqrt{1+4\xi_3}}{2}\right)^2+\xi_4\right],$$

which alternatively becomes

$$E_{n\kappa}^2=m_0^2-\frac{4\left(m_0-E_{n\kappa}\right)^2V_2^2}{\alpha^2\left(1+2n+\sqrt{\left(2\kappa+2T-1\right)^2-\frac{8\left(m_0-E_{n\kappa}\right)V_1}{\alpha^2}}\right)^2}$$
$$-\frac{\alpha^2\left(1+2n+\sqrt{\left(2\kappa+2T-1\right)^2-\frac{8\left(m_0-E_{n\kappa}\right)V_1}{\alpha^2}}\right)^2}{4}, \tag{32}$$

where we used the transformations

$$E_{n\kappa}\to-E_{n\kappa};\ V(r)\to-V(r)\ \text{(i.e. } V_1\to-V_1,\ V_2\to-V_2);\ \kappa\to\kappa-1;$$
$$F_{n\kappa}(r)\to G_{n\kappa}(r);\ \text{(or } \xi_1\to\xi_1 \text{ and } \xi_2\to\xi_4). \tag{33}$$

The lower spinor component of the wave function can be obtained as

$$G_{n\kappa}(r)=N''\left(e^{-2\alpha r}\right)^{\sqrt{\xi_0-\xi_4}}(1-e^{-2\alpha r})^{\left(1+\sqrt{1+4\xi_3}\right)/2}$$
$$\times{}_2F_1\left(-n,n+2\sqrt{\xi_0-\xi_4}+\sqrt{1+4\xi_3}+1;2\sqrt{\xi_0-\xi_4}+1;e^{-2\alpha r}\right),$$
$$N''=\frac{\Gamma\left(n+2\sqrt{\xi_0-\xi_4}+1\right)}{\Gamma\left(2\sqrt{\xi_0-\xi_4}+1\right)n!}N, \tag{34}$$

where $N''$ is the normalization constant. Finally, the upper-spinor component of the Dirac equation is found via Eq. (9b) as

$$F_{n\kappa}(r)=\frac{1}{\left(m_0-E_{n\kappa}\right)}\left\{\left[\alpha\left(1+\sqrt{1+4\xi_3}\right)\frac{e^{-2\alpha r}}{(1-e^{-2\alpha r})}-2\alpha\sqrt{\xi_0-\xi_4}+\frac{\kappa+T-1}{r}\right]G_{n\kappa}(r)\right.$$
$$+N''\left(e^{-2\alpha r}\right)^{1+\sqrt{\xi_0-\xi_4}}(1-e^{-2\alpha r})^{\left(1+\sqrt{1+4\xi_3}\right)/2}\frac{2\alpha n\left(n+2\sqrt{\xi_0-\xi_4}+\sqrt{1+4\xi_3}+1\right)}{\left(2\sqrt{\xi_0-\xi_4}+1\right)} \tag{35}$$
$$\left.\times{}_2F_1\left(1-n,n+2\sqrt{\xi_0-\xi_4}+\sqrt{1+4\xi_3}+2;2\sqrt{\xi_0-\xi_4}+2;e^{-2\alpha r}\right)\right\}.$$

## 3. Discussions

In this section in the framework of the Dirac theory with the Eckart potential, we obtain the energy equation and spinor wave functions for several well-known potentials by choosing appropriate parameters in the Eckart potential model.

### 3.1. Hulthén potential

If we take $\alpha=\frac{1}{2a}$, $V_1=0$ and $V_2=\frac{V_0}{2}$, the Eckart potential in Eq. (1) turns to the standard Hulthén potential with a constant shift of $\frac{V_0}{2}$ [63]

$$V(r)=-V_0\frac{e^{-2\alpha r}}{1-e^{-2\alpha r}}-\frac{V_0}{2}. \tag{36}$$

The Hulthén potential is widely used in atomic physics [64,65], solid state physics [66] and chemical physics [67]. Using Eq. (17) and making the corresponding parameter replacements in Eq. (23), we obtain the energy equation for the Hulthén potential in the Dirac theory as

$$m_0^2-E_{n\kappa}^2=\left[\frac{aV_0\left(m_0+E_{n\kappa}\right)}{n+\kappa+T+1}-\frac{n+\kappa+T+1}{2a}\right]^2. \tag{37}$$

Hence, making the appropriate replacements, we consequently obtain the wave functions $F_{n\kappa}(r)$ and $G_{n\kappa}(r)$ as

$$F_{n\kappa}(r)=Ne^{-\sqrt{m_0^2-E_{n\kappa}^2}r}(1-e^{-r/a})^{\kappa+T+1}P_n^{\left(2a\sqrt{m_0^2-E_{n\kappa}^2},2\kappa+2T+1\right)}\left(1-2e^{-r/a}\right),$$
$$=\mathrm{N}'e^{-\sqrt{m_0^2-E_{n\kappa}^2}r}(1-e^{-r/a})^{\kappa+T+1}$$
$$\times{}_2F_1\left(-n,n+2a\sqrt{m_0^2-E_{n\kappa}^2}+2(\kappa+T+1);2a\sqrt{m_0^2-E_{n\kappa}^2}+1;e^{-r/a}\right),\tag{38}$$
$$\mathrm{N}'=\frac{\Gamma\left(n+2a\sqrt{m_0^2-E_{n\kappa}^2}+1\right)}{\Gamma\left(2a\sqrt{m_0^2-E_{n\kappa}^2}+1\right)n!}N,\ \sqrt{m_0^2-E_{n\kappa}^2}=\frac{aV_0(m_0+E_{n\kappa})}{n+\kappa+T+1}-\frac{n+\kappa+T+1}{2a},$$

and

$$G_{n\kappa}(r)=\frac{1}{(m_0+E_{n\kappa})}\left\{\left[\frac{(\kappa+T+1)}{a}\frac{e^{-r/a}}{(1-e^{-r/a})}-\sqrt{m_0^2-E_{n\kappa}^2}+\frac{\kappa+T}{r}\right]F_{n\kappa}(r)\right.$$
$$+\mathrm{N}'e^{-\sqrt{m_0^2-E_{n\kappa}^2}r}e^{-r/a}(1-e^{-r/a})^{\kappa+T+1}\frac{n\left(n+2a\sqrt{m_0^2-E_{n\kappa}^2}+2(\kappa+T+1)\right)}{a\left(2a\sqrt{m_0^2-E_{n\kappa}^2}+1\right)}\tag{39}$$
$$\left.\times{}_2F_1\left(1-n,n+n+2a\sqrt{m_0^2-E_{n\kappa}^2}+2(\kappa+T+1)+1;2a\sqrt{m_0^2-E_{n\kappa}^2}+2;e^{-r/a}\right)\right\}.$$

These results are consistent with those given in Eqs. (50) and (55) of Ref. [59].

### 3.2. Generalized Morse potential

The generalized Morse potential (GMP) is used to describe diatomic molecular energy spectra and electromagnetic transitions. It is defined by [68]

$$V(r)=D\left[1-\frac{b}{e^{ar}-1}\right]^2,\tag{40}$$

with $b=e^{ar_0}-1$, and $D$, $b$, $a$ are some parameters regulating the depth, position of the minimum $r_0$, and radius of the potential. Codriansky et al. [69] showed that the solvability of the GMP is due to the fact that it belongs to the class of the Eckart potential, a member of the hypergeometric Natanzon potentials, which can be solved algebraically by means of $SO(2,1)$ algebra. If we take $\alpha=\frac{a}{2}$, $V_1=\frac{1}{4}b^2D$ and $V_2=bD\left(1+\frac{1}{2}b\right)$, the Eckart potential in Eq. (1) turns to the GMP with a constant shift

$$V(r)=D\left[1-\frac{b}{e^{ar}-1}\right]^2-D\left(1+b+\frac{1}{2}b^2\right).\tag{41}$$

Using Eq. (17) and making the corresponding parameter replacements in Eq. (23), we obtain the energy equation for the GMP in the Dirac theory as

$$m_0^2 - E_{n\kappa}^2 = \left( \frac{4\left(m_0 + E_{n\kappa}\right) bD\left(1 + \frac{1}{2}b\right)}{a\left(2n + \eta_\kappa\right)} - \frac{a\left(2n + \eta_\kappa\right)}{4} \right)^2 - 2D\left(m_0 + E_{n\kappa}\right), \tag{42}$$

where $\eta_\kappa = 1 + \sqrt{\left(2\kappa + 2T + 1\right)^2 + \frac{8\left(m_0 + E_{n\kappa}\right) b^2 D}{a^2}}$. Making the convenient replacements, we consequently obtain the wave functions $F_{n\kappa}(r)$ and $G_{n\kappa}(r)$ as

$$\begin{aligned} F_{n\kappa}(r) &= N e^{-\sqrt{m_0^2 - E_{n\kappa}^2}\, r} (1 - e^{-ar})^{\eta_\kappa/2} P_n^{\left(\frac{2}{a}\sqrt{m_0^2 - E_{n\kappa}^2},\, \eta_\kappa - 1\right)} \left(1 - 2e^{-ar}\right), \\ &= \mathrm{N}' e^{-\sqrt{m_0^2 - E_{n\kappa}^2}\, r} (1 - e^{-ar})^{\eta_\kappa/2} {}_2F_1\left(-n, n + \frac{2}{a}\sqrt{m_0^2 - E_{n\kappa}^2} + \eta_\kappa; \frac{2}{a}\sqrt{m_0^2 - E_{n\kappa}^2} + 1; e^{-ar}\right), \end{aligned} \tag{43}$$

where

$$\begin{aligned} \mathrm{N}' &= \frac{\Gamma\left(n + \frac{2}{a}\sqrt{m_0^2 - E_{n\kappa}^2} + 1\right)}{\Gamma\left(\frac{2}{a}\sqrt{m_0^2 - E_{n\kappa}^2} + 1\right) n!} N, \\ \sqrt{m_0^2 - E_{n\kappa}^2} &= \sqrt{\left( \frac{4\left(m_0 + E_{n\kappa}\right) bD\left(1 + \frac{1}{2}b\right)}{a\left(2n + \eta_\kappa\right)} - \frac{a\left(2n + \eta_\kappa\right)}{4} \right)^2 - 2D\left(m_0 + E_{n\kappa}\right)}, \end{aligned} \tag{44}$$

and

$$\begin{aligned} G_{n\kappa}(r) = \frac{1}{\left(m_0 + E_{n\kappa}\right)} &\left\{ \left[ \frac{a\eta_\kappa}{2} \frac{e^{-ar}}{(1 - e^{-ar})} - \sqrt{m_0^2 - E_{n\kappa}^2} + \frac{\kappa + T}{r} \right] F_{n\kappa}(r) \right. \\ &+ \mathrm{N}' \frac{a^2 n\left(n + \frac{2}{a}\sqrt{m_0^2 - E_{n\kappa}^2} + \eta_\kappa\right)}{2\sqrt{m_0^2 - E_{n\kappa}^2} + 1} e^{-\sqrt{m_0^2 - E_{n\kappa}^2}\, r} (1 - e^{-ar})^{\eta_\kappa/2} e^{-ar} \\ &\left. \times {}_2F_1\left(1 - n, n + \frac{2}{a}\sqrt{m_0^2 - E_{n\kappa}^2} + \eta_\kappa + 1; \frac{2}{a}\sqrt{m_0^2 - E_{n\kappa}^2} + 2; e^{-ar}\right) \right\}. \end{aligned} \tag{45}$$

### 3.3. The nonrelativistic Eckart potential

In the non-relativistic limit, the energy levels can be obtained when $T = 0$, $E_{n\kappa} + m_0 \approx 2\mu/\hbar^2$, $E_{n\kappa} - m_0 \approx E_{nl}$ and $\kappa(\kappa + 1) \rightarrow l(l + 1)$ from (23) as

$$E_{nl}=-\frac{2\mu V_2^2}{\alpha^2\hbar^2\left(n+1/2+\sqrt{(l+1/2)^2+\frac{4\mu V_1}{\alpha^2\hbar^2}}\right)^2}$$
$$-\frac{\alpha^2\hbar^2}{2\mu}\left(n+1/2+\sqrt{(l+1/2)^2+\frac{4\mu V_1}{\alpha^2\hbar^2}}\right)^2, \tag{46}$$

and the radial wave functions from Eq. (28) are

$$R_{nl}(r)=\mathrm{N}'\left(e^{-2\alpha r}\right)^{\lambda}(1-e^{-2\alpha r})^{(1+\sigma)/2}\,{}_2F_1\left(-n,n+2\lambda+\sigma+1;2\lambda+1;e^{-2\alpha r}\right),$$
$$\mathrm{N}'=\frac{\Gamma(n+2\lambda+1)}{\Gamma(2\lambda+1)n!}N,\ \lambda=\sqrt{-\frac{\mu(E_{nl}+2V_2)}{2\alpha^2\hbar^2}},\ \sigma=2\sqrt{(l+1/2)^2+\frac{4\mu V_1}{\alpha^2\hbar^2}}. \tag{47}$$

### 3.4. Solution via Asymptotic Iteration Method

By using the asymptotic iteration method (AIM), Bahar and Yasuk [70] presented approximate solutions of the Dirac equation with the Eckart potential in the case of PDM. In this subsection, we show that their presented analytical energy formula and consequently their numerical results are incorrect. Equation (24) of [70] should be rewritten in a more simple form as

$$f''(s)+\left[\frac{(2p+1)-(2p+q+1)s}{s(1-s)}\right]f'(s)-\left[\frac{(p+q)^2-\xi_1-\xi_2}{s(1-s)}\right]f(s)=0, \tag{48}$$

where $p=\sqrt{\xi_0-\xi_2}$ and $q=\frac{1}{2}\left[1+\sqrt{1+4\xi_1}\right]$.

As a consequence, equation (25) in [70] should be

$$\lambda_0(s)=\left[\frac{(2p+q+1)s-(2p+1)}{s(1-s)}\right],\qquad s_0(s)=\left[\frac{(p+q)^2-\xi_1-\xi_2}{s(1-s)}\right],$$
$$\lambda_1(s)=\frac{1}{s^2(1-s)^2}\Big\{\Big(2p^3(-1+s)-q^2(1-s)^2+(1+\xi_0+\xi_2)(1-s)^2+2q^3s+2qs(s-\xi_0-\xi_2)\Big)$$
$$+2p\Big(-q(1-s)^2+(1-s)(s-1-\xi_0-\xi_2)+q^2(3s-1)\Big)+p^2\Big(-(1-s)^2+q(6s-4)\Big)\Big\},$$
$$s_1(s)=\frac{\left(p^2+2\mathrm{pq}+q^2-\xi_0-\xi_2\right)(2p(s-1)-2+(3+2q)s)}{s^2(1-s)^2}, \tag{49}$$

Thus, combining the results obtained by means of AIM with the quantization condition given by equation. (21) of Ref. [70] yields:

$$s_0\lambda_1-s_1\lambda_0=0\Rightarrow p+q=0+\sqrt{\xi_0+\xi_2},\qquad \text{for}\quad n=0,$$
$$s_0\lambda_1-s_1\lambda_0=0\Rightarrow p+q=-1+\sqrt{\xi_0+\xi_2},\qquad \text{for}\quad n=1, \tag{50}$$
$$s_0\lambda_1-s_1\lambda_0=0\Rightarrow p+q=-2+\sqrt{\xi_0+\xi_2},\qquad \text{for}\quad n=2,$$
$$\vdots$$

Therefore, when the above expressions are generalized, the eigenvalues turn out to become

$$\xi_0+\xi_2=\left[\frac{2\xi_2}{\left(2n+1+\sqrt{1+4\xi_1}\right)}+\frac{\left(2n+1+\sqrt{1+4\xi_1}\right)}{4}\right]^2, \tag{51}$$

Now, upon using equation. (27) of Ref. [70], we give the correct Dirac energy states, in the framework of AIM, for this potential as

$$\left(m_0+E\right)\left(m_0-E+2V_2\right)=4\alpha^2\left[\frac{\frac{4\left(m_0+E\right)V_2}{\alpha^2}+\left(2n+1+\sqrt{\left(2\kappa+1\right)^2+\frac{8V_1(m_0+E)}{\alpha^2}}\right)^2}{4\left(2n+1+\sqrt{\left(2\kappa+1\right)^2+\frac{8V_1(m_0+E)}{\alpha^2}}\right)}\right]^2. \tag{52}$$

Notice that Eq. (52) is exactly the same as equation (23) which we have already obtained by using pNU when $T=0$. We have thus shown the correct numerical results in table 1.

**4. Conclusion**

To sum up, in this paper, we have investigated the analytical approximate bound state solutions of the position-dependent mass Dirac equation with the vector Eckart potential including Coulomb-like tensor interaction. Further, we have used a suitable mass function and considered a usual approximation scheme to deal with the centrifugal kinetic term in the framework of the pNU method. The energy eigenvalues equation and the corresponding normalized two-spinor components of the wave functions are calculated in a systematic way for any arbitrary spin-orbit quantum number $\kappa$ in the low screening regime where $\alpha$ is small. The eigensolutions of some special cases like the Hulthén potential and the generalized Morse potential are also given.

To show the accuracy of the present model, some numerical values of the energy levels are calculated in table 1 for the lowest states. Obviously, the degeneracy between the members of doublet states in spin and p-spin symmetries is removed by tensor interaction. Finally, the relativistic spin symmetry in the absence of tensor interaction ($T=0$) reduces into the Schrödinger solution for the Eckart potential under appropriate transformations of parameters.

Finally, we have found that Dirac energy equations, Eqs. (23) and (32), with spatially-dependent mass particle subjected to the vector Eckart potential field and the Coulomb tensor interaction potential are found to be same as the Dirac energy equations with constant mass particle in the field of equally mixed scalar-vector

Eckart potentials including Coulomb tensor interaction under spin $\left(\Sigma = 2V_{\text{Eckart}}(r),\ \Delta(r) = 0\right)$ and pseudospin $\left(\Sigma = 0,\ \Delta(r) = 2V_{\text{Eckart}}(r)\right)$ symmetries, respectively. The numerical Dirac energy bound states in presence of vector Eckart potential are calculated within the pNU method and AIM in the presence of spin and p-spin symmetries as shown in tables 1 and 2, respectively.

**Acknowledgments**

The author would like to thank the kind referees for the useful comments and suggestions which have improved the manuscript greatly. BJ Falaye dedicates this work to his parents for their love.

**Appendix A: A Review to Parametric Nikiforov-Uvarov Method**

This powerful mathematical tool solves second order differential equations. Let us consider the following differential equation

$$\left[\frac{d^2}{ds^2} + \frac{\tilde{\tau}(s)}{\sigma(s)}\frac{d}{ds} + \frac{\tilde{\sigma}(s)}{\sigma^2(s)}\right]\psi_n(s) = 0, \tag{A1}$$

where $\sigma(s)$ and $\tilde{\sigma}(s)$ are polynomials, at most of second degree, and $\tilde{\tau}(s)$ is a first-degree polynomial. To make the application of the NU method [30] simpler and direct without need to check the validity of solution. At first we write the general form of the Schrödinger-like equation (A1) in a more general form as

$$\left[\frac{d^2}{ds^2} + \frac{c_1 - c_2 s}{s(1-c_3 s)}\frac{d}{ds} + \frac{1}{s^2(1-c_3 s)^2}\left(-p_2 s^2 + p_1 s - p_0\right)\right]\psi_n(s) = 0, \tag{A2}$$

satisfying the wave functions

$$\psi_n(s) = \phi(s) y_n(s). \tag{A3}$$

Comparing (A2) with its counterpart (A1), we obtain the following identifications:

$$\tilde{\tau}(s) = c_1 - c_2 s, \quad \sigma(s) = s(1-c_3 s), \quad \tilde{\sigma}(s) = -p_2 s^2 + p_1 s - p_0, \tag{A4}$$

(1) For the given choice of root $k_-$ and the function $\pi(s)$:

$$k_- = -(c_7 + 2c_3 c_8) - 2\sqrt{c_8 c_9}, \quad \pi(s) = c_4 + \sqrt{c_8} - \left(\sqrt{c_9} + c_3\sqrt{c_8} - c_5\right)s,$$

we follow the NU method [30] to obtain the energy equation [31]

$$nc_2 - (2n+1)c_5 + (2n+1)\left(\sqrt{c_9} + c_3\sqrt{c_8}\right) + n(n-1)c_3 + c_7 + 2c_3c_8 + 2\sqrt{c_8c_9} = 0, \tag{A5}$$

and the wave functions

$$\rho(s) = s^{c_{10}}(1-c_3 s)^{c_{11}}, \quad \phi(s) = s^{c_{12}}(1-c_3 s)^{c_{13}}, \ c_{12} > 0, \ c_{13} > 0,$$

$$y_n(s) = P_n^{(c_{10},c_{11})}(1-2c_3 s), \ c_{10} > -1, \ c_{11} > -1,$$

$$\psi_{n\kappa}(s) = N_{n\kappa} s^{c_{12}}(1-c_3 s)^{c_{13}} P_n^{(c_{10},c_{11})}(1-2c_3 s), \tag{A6}$$

where $P_n^{(\mu,\nu)}(x), \ \mu > -1, \ \nu > -1$, and $x \in [-1,1]$ are Jacobi polynomials with the constants are [31-35]

$$c_4 = \frac{1}{2}(1-c_1), \qquad c_5 = \frac{1}{2}(c_2 - 2c_3),$$

$$c_6 = c_5^2 + p_2; \qquad c_7 = 2c_4 c_5 - p_1,$$

$$c_8 = c_4^2 + p_0, \qquad c_9 = c_3(c_7 + c_3 c_8) + c_6,$$

$$c_{10} = 2\sqrt{c_8} > -1, \qquad c_{11} = \frac{2}{c_3}\sqrt{c_9} > -1, \ c_3 \neq 0,$$

$$c_{12} = c_4 + \sqrt{c_8} > 0, \qquad c_{13} = -c_4 + \frac{1}{c_3}(\sqrt{c_9} - c_5) > 0, \ c_3 \neq 0, \tag{A7}$$

where $c_{12} > 0, \ c_{13} > 0$ and $s \in [0, 1/c_3], \ c_3 \neq 0.$

In the rather more special case of $c_3 = 0$, the wave function (A3) becomes

$$\lim_{c_3 \to 0} P_n^{(c_{10},c_{11})}(1-2c_3 s) = L_n^{c_{10}}\left(2\sqrt{c_9}\, s\right), \ \lim_{c_3 \to 0}(1-c_3 s)^{c_{13}} = e^{-(\sqrt{c_9} - c_5)s},$$

$$\psi(s) = N s^{c_{12}} e^{-(\sqrt{c_9} - c_5)s} L_n^{c_{10}}(2\sqrt{c_9}\, s). \tag{A8}$$

(2) For the given root $k_+$ and the function $\pi(s)$:

$$k_+ = -(c_7 + 2c_3 c_8) + 2\sqrt{c_8 c_9}, \ \pi(s) = c_4 - \sqrt{c_8} - \left(\sqrt{c_9} - c_3\sqrt{c_8} - c_5\right)s,$$

we follow the NU method [30] to obtain the energy equation

$$nc_2 - (2n+1)c_5 + (2n+1)\left(\sqrt{c_9} - c_3\sqrt{c_8}\right) + n(n-1)c_3 + c_7 + 2c_3 c_8 - 2\sqrt{c_8 c_9} = 0, \tag{A9}$$

and the wave functions

$$\rho(s) = s^{\tilde{c}_{10}}(1-c_3 s)^{\tilde{c}_{11}}, \quad \phi(s) = s^{\tilde{c}_{12}}(1-c_3 s)^{\tilde{c}_{13}}, \ \tilde{c}_{12} > 0, \ \tilde{c}_{13} > 0,$$

$$y_n(s) = P_n^{(\tilde{c}_{10},\tilde{c}_{11})}(1-2\tilde{c}_3 s), \ \tilde{c}_{10} > -1, \ \tilde{c}_{11} > -1,$$

$$\psi_{n\kappa}(s) = N_{n\kappa} s^{\tilde{c}_{12}}(1-c_3 s)^{\tilde{c}_{13}} P_n^{(\tilde{c}_{10},\tilde{c}_{11})}(1-2c_3 s), \tag{A10}$$

with

$$\tilde{c}_{10} = -2\sqrt{c_8}, \qquad \tilde{c}_{11} = \frac{2}{c_3}\sqrt{c_9},\ c_3 \neq 0,$$

$$\tilde{c}_{12} = c_4 - \sqrt{c_8} > 0, \qquad \tilde{c}_{13} = -c_4 + \frac{1}{c_3}(\sqrt{c_9} - c_5) > 0,\ c_3 \neq 0. \tag{A11}$$

**Appendix B: Relativistic Normalization Constant**

Unlike the nonrelativistic case, the normalization condition for the Dirac spinor combines the two individual normalization constants in a single integral. The radial wave functions are normalized according to the formula [56]

$$\int_0^\infty \psi^\dagger(r)\psi(r)d^3r = \int_0^\infty \left(|F_{n\kappa}(r)|^2 + |G_{n\kappa}(r)|^2\right)dr = 1, \tag{B1}$$

and can be rewritten in terms of the new variable $s = e^{-2\alpha r}$ as

$$\int_0^1 \left(|F_{n\kappa}(s)|^2 + |G_{n\kappa}(s)|^2\right)\frac{ds}{s} = 2\alpha. \tag{B2}$$

Therefore, substituting Eqs. (28) and (34), we can obtain

$$N^2\left[\left(\frac{\Gamma(n+2b+1)}{\Gamma(2b+1)n!}\right)^2 \int_0^1 s^{2b-1}(1-s)^{a+1}\left[{}_2F_1(-n, n+2b+a+1; 2b+1; s)\right]^2 ds\right.$$
$$\left. + \left(\frac{\Gamma(n+2c+1)}{\Gamma(2c+1)n!}\right)^2 \int_0^1 s^{2c-1}(1-s)^{d+1}\left[{}_2F_1(-n, n+2c+d+1; 2c+1; s)\right]^2 ds\right] = 2\alpha, \tag{B3}$$

where $b = \sqrt{\xi_0 - \xi_2}$, $c = \sqrt{\xi_0 - \xi_4}$, $a = \sqrt{1+4\xi_1}$ and $d = \sqrt{1+4\xi_3}$.

Then we obtain the normalization constant as [71,72]

$$N = \sqrt{2\alpha}\left[\left(\frac{\Gamma(n+2b+1)}{\Gamma(2b+1)n!}\right)^2 \sum_{j=0}^{\infty} \frac{(-n)_j(n+2b+a+1)_j(2b)_j}{(2b+1)_j(2b++a+2)_j\, j!} B(2b, a+2)\right.$$
$$\times {}_3F_2(-n, n+2b+a+1; 2b+j; 2b+1, 2b+a+2+j; s)$$
$$+ \left(\frac{\Gamma(n+2c+1)}{\Gamma(2c+1)n!}\right)^2 \sum_{j=0}^{\infty} \frac{(-n)_j(n+2c+d+1)_j(2c)_j}{(2c+1)_j(2c+d+2)_j\, j!} B(2c, d+2)$$
$$\left. \times {}_3F_2(-n, n+2c+d+1; 2c+j; 2c+1, 2c+d+2+j; s)\right]^{-1/2}, \tag{B4}$$

with the confluent hypergeometric function defined by

$$ {}_pF_q\left(a_1,\ldots,a_p;b_1,\ldots,b_q;s\right)=\sum_{j=0}^{\infty}\frac{(a_1)_j\ldots(a_p)_j s^j}{(b_1)_j\ldots(b_q)_j\, j!}, \tag{B5}$$

where $(a_1)_{j,}$ $(b_1)_j$ are Pochhammer symbols. Overmore, the beta function is defined by

$$B(x,y)=B(y,x)=\int_0^1 t^{x-1}\left(1-t\right)^{y-1}dt=\frac{\Gamma(x)\Gamma(y)}{\Gamma(x+y)},\ \ \mathrm{Re}(x),\ \ \mathrm{Re}(y)>0, \tag{B6}$$

with $B(1/2,1/2)=\pi.$

The incomplete beta function is defined by

$$B(x;a,b)=\int_0^x t^{a-1}\left(1-t\right)^{b-1}dt, \tag{B7}$$

and the regularized beta function is

$$I_x(a,b)=\frac{B(x;a,b)}{B(a,b)}. \tag{B8}$$

**Table 1.** Bound state energy states (in atomic units $\hbar = m_0 = 1$) in the spatially-dependent mass spin symmetric Dirac equation with the vector Eckart potential including tensor interaction.

| $V_1 = V_2 = 0.8$ | | | | $V_1 = V_2 = 1.0$ | | | |
|---|---|---|---|---|---|---|---|
| $\alpha$ | $\|n,l,\kappa\rangle$ | $E_{n\kappa}$ (T = 0) AIM & pNU | $E_{n\kappa}$ (T = 0.5) | $\alpha$ | $n,l,\kappa$ | $E_{n\kappa}$ (T = 0) AIM & pNU | $E_{n\kappa}$ (T = 0.5) |
| 0.005 | $\|0,0,0\rangle$ | -1.061765390 | -1.058238387 | 0.005 | $\|0,0,0\rangle$ | -1.037763395 | -1.499707182 |
| 0.005 | $\|1,0,0\rangle$ | -1.126296185 | -1.124245136 | 0.005 | $\|1,0,0\rangle$ | -1.524320061 | -1.522125986 |
| 0.005 | $\|2,1,1\rangle$ | -1.170316259 | -1.166862392 | 0.005 | $\|2,1,1\rangle$ | -1.550079490 | -1.546020820 |
| 0.005 | $\|3,2,2\rangle$ | -1.205260185 | -1.200914621 | 0.005 | $\|3,2,2\rangle$ | -1.575237513 | -1.570025662 |
| 0.1 | $\|0,0,0\rangle$ | -1.094849742 | -1.086083590 | 0.1 | $\|0,0,0\rangle$ | -1.509062425 | -1.499037037 |
| 0.1 | $\|1,0,0\rangle$ | -1.194138407 | -1.188639607 | 0.1 | $\|1,0,0\rangle$ | -1.569136849 | -1.562840954 |
| 0.1 | $\|2,1,1\rangle$ | -1.256811175 | -1.247711802 | 0.1 | $\|2,1,1\rangle$ | -1.620269182 | -1.609727882 |
| 0.1 | $\|3,2,2\rangle$ | -1.303836766 | -1.292645011 | 0.1 | $\|3,2,2\rangle$ | -1.662494255 | -1.649606042 |
| 0.3 | $\|0,0,0\rangle$ | -1.180089379 | -1.144590267 | 0.3 | $\|0,0,0\rangle$ | -1.552435590 | -1.499557382 |
| 0.3 | $\|1,0,0\rangle$ | -1.378977823 | -1.352933326 | 0.3 | $\|1,0,0\rangle$ | -1.738803862 | -1.710886318 |
| 0.3 | $\|2,1,1\rangle$ | -1.474676212 | -1.435078704 | 0.3 | $\|2,1,1\rangle$ | -1.836358965 | -1.793697389 |
| 0.3 | $\|3,2,2\rangle$ | -1.530829536 | -1.486501313 | 0.3 | $\|3,2,2\rangle$ | -1.897332069 | -1.849115091 |

**Table 2.** Bound state energy levels (in atomic units $\hbar = \mu = m_0 = 1$) in the spatially-dependent mass pseudospin symmetric Dirac equation with the Eckart potential including tensor interaction.

| $V_1 = V_2 = 0.8$ | | | | $V_1 = V_2 = 1.0$ | | | |
|---|---|---|---|---|---|---|---|
| $\alpha$ | $\|n,l,\kappa\rangle$ | $E_{n\kappa}$ (T = 0) pNU | $E_{n\kappa}$ (T = 0.5) pNU | $\alpha$ | $n,l,\kappa$ | $E_{n\kappa}$ (T = 0) pNU | $E_{n\kappa}$ (T = 0.5) pNU |
| 0.005 | $\|0,0,0\rangle$ | -2.999679672 | -2.999796918 | 0.005 | $\|0,0,0\rangle$ | -3.499722214 0.9724876465 | -3.499826419 |
| 0.005 | $\|1,0,0\rangle$ | -2.998057613 | -2.998175328 | 0.005 | $\|1,0,0\rangle$ | -3.498335419 | -3.498439931 |
| 0.005 | $\|2,1,1\rangle$ | -2.994829867 | -2.994473928 | 0.005 | $\|2,1,1\rangle$ | -3.495573359 | -3.495257975 |
| 0.005 | $\|3,2,2\rangle$ | -2.989068945 | -2.988228670 | 0.005 | $\|3,2,2\rangle$ | -3.490610772 | -3.489868420 |
| 0.1 | $\|0,0,0\rangle$ | -2.998718500 | -2.999188181 | 0.1 | $\|0,0,0\rangle$ | -3.498888752 0.9724876465 | -3.499306036 |
| 0.1 | $\|1,0,0\rangle$ | -2.992265612 | -2.992742750 | 0.1 | $\|1,0,0\rangle$ | -3.493366728 | -3.493788925 |
| 0.1 | $\|2,1,1\rangle$ | -2.979622680 | -2.978147160 | 0.1 | $\|2,1,1\rangle$ | -3.482507804 | -3.481212039 |
| 0.1 | $\|3,2,2\rangle$ | -2.957210583 | -2.953614128 | 0.1 | $\|3,2,2\rangle$ | -3.463114139 | -3.459988846 |
| 0.3 | $\|0,0,0\rangle$ | -2.988448520 | -2.992742750 | 0.3 | $\|0,0,0\rangle$ | -3.489988927 0.9021648060 | -3.493788925 |
| 0.3 | $\|1,0,0\rangle$ | -2.933817147 | -2.938681640 | 0.3 | $\|1,0,0\rangle$ | -3.442738552 | -3.446918383 |
| 0.3 | $\|2,1,1\rangle$ | -2.845698202 | -2.828296539 | 0.3 | $\|2,1,1\rangle$ | -3.363288460 | -3.348791462 |
| 0.3 | $\|3,2,2\rangle$ | -2.706762262 | -2.658840360 | 0.3 | $\|3,2,2\rangle$ | -3.234826441 | -3.195484406 |

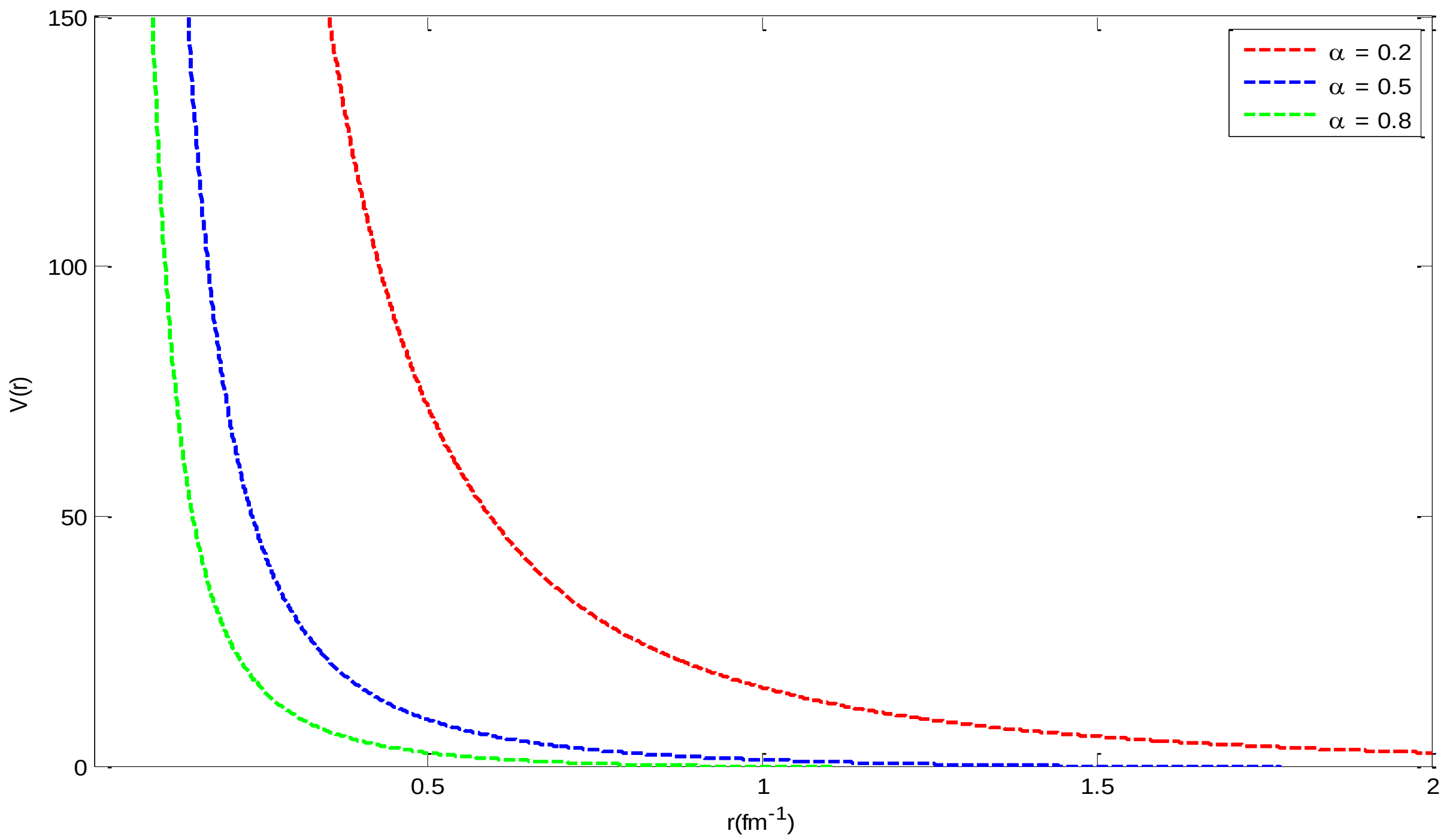

Figure 1. The behavior of the Eckart potential with $r$ for different values of screening parameter $\alpha$ and taking $V_1 = V_2 = 0.8$.